# Strong-coupling and high-bandwidth cavity electro-optic modulation for advanced pulse-comb synthesis

Tianqi Lei[1,*], Yunxiang Song[2,3,*], Yanyun Xue[1], Qihuang Gong[1,4], Marko Lončar[2,†], Yaowen Hu[1,4,†]

[1]State Key Laboratory for Mesoscopic Physics and Frontiers Science Center for Nano-optoelectronics, School of Physics, Peking University, Beijing 100871, China

[2]John A. Paulson School of Engineering and Applied Sciences, Harvard University, Cambridge, MA 02138, USA

[3]Quantum Science and Engineering, Harvard University, Cambridge, MA 02138, USA

[4]Collaborative Innovation Center of Extreme Optics, Shanxi University, Taiyuan, China.

[*]These authors contributed equally.

[†]Corresponding authors: loncar@seas.harvard.edu; yaowenhu@pku.edu.cn;

## Abstract

Cavity electro-optic (EO) modulation plays a pivotal role in optical pulse and frequency comb synthesis, supporting a wide range of applications including communication, computing, ranging, and quantum information. The ever-growing demand for these applications has driven efforts in enhancing modulation coupling strength and bandwidth towards advanced pulse-comb synthesis. However, the effects of strong-coupling and high-bandwidth cavity EO modulation remain underexplored, due to the lack of a general, unified model that captures this extreme condition. In this work, we present a universal framework for pulse-comb synthesis under cavity EO modulation, where coupling strength and modulation bandwidth far exceed the cavity's free spectral range (FSR). We show that, under such intense and ultrafast driving conditions, EO-driven frequency combs and pulses exhibit rich higher-order nonlinear dynamics, including temporal pulse compression and comb generation with arbitrary pump detuning. Leveraging this framework, we reveal a direct link between the higher-order dynamics of EO pulse-comb generation and the band structure of synthetic dimension. Furthermore, we demonstrate arbitrary comb shaping via machine-learning-based inverse microwave drive design, achieving a tenfold enhancement in cavity electro-optic comb flatness by exploring the synergistic effects of high-bandwidth driving and detuning-induced frequency boundaries. Our findings push cavity electro-optic modulation into a new frontier, unlocking significant potential for universal and machine-learning-programmable electro-optic frequency combs, topological photonics, as well as photonic quantum computing in the strong-coupling and high-bandwidth regimes.

## Introduction

Cavity electro-optic (EO) modulation is a foundational mechanism in modern optoelectronics, enabling the rapid frequency tuning of optical cavities to generate ultrafast optical pulses[1,2] and coherent frequency combs[3–6]. It provides a stable and reconfigurable approach to pulse-comb synthesis with a direct interface to high-speed electronics. This technique has found widespread application in diverse fields such as modern communications[7–10], photonic computing[11,12], topological photonics[13,14], synthetic crystal generation[14–16], precision measurement and ranging[17,18], and quantum information science[19–22]. In recent years, there has been a growing demand for enhanced EO modulation capabilities, specifically stronger coupling strengths and broader modulation bandwidths, to meet the critical requirements of emerging advanced photonic technologies[23]. These demands include expanding the EO comb span, reducing temporal pulse durations, improving energy efficiency, and boosting tuning speed for communications, computing, ranging, etc. Despite these advancements, the key physical process — cavity EO modulation under strong microwave-driven coupling between photonic cavity modes and high-frequency driving bandwidth — remains largely unexplored. Especially, from a physics point of view, the dynamics of cavity electro-optics for related EO comb/pulse system, has received less attention compared to the recent cavity $\chi^{(3)}$nonlinearity for Kerr comb/soliton generation[24,25], mainly due to the apparent simplicity of cavity EO modulation[4,5]. However, there still exists a largely underexplored regime of electro-optics that may exhibit complex, rich physics due to the nonlinear nature of cavity EO modulation response to the microwave drive strength.

Conventional cavity electro-optic (EO) modulation typically achieves an EO coupling strength $\Omega$ (or equivalently, the modulation index $\beta = 2\pi \frac{\Omega}{\omega_{\mathrm{R}}}$, where $\omega_{\mathrm{R}}$ is the cavity's free spectral range (FSR))[5] that is comparable to/larger than the cavity's resonant linewidth $\kappa$ (sideband unresolved/resolved regime[23]). It typically uses a microwave drive with bandwidth $\omega_{\mathrm{BW}}$ on the order of $\kappa$ for communications applications[26,27], or with $\omega_{\mathrm{BW}}$ equal to $\omega_{\mathrm{R}}$ for efficient frequency comb and optical pulse generation[3–5]. Under these conditions, cavity EO modulation operates within a phase space defined by the coupling strength $\Omega$ and modulation bandwidth $\omega_{\mathrm{BW}}$, typically constrained to within an area of $\omega_{\mathrm{R}} \times \omega_{\mathrm{R}}$ (Fig. 1a). Though the physics in this regime has a relatively simple mechanism, it has already stimulated various advanced EO devices and applications in recent years, such as cavity EO comb generation[4,5], EO frequency shifter and single sideband modulation[28], photonic molecule and isolation[29,30], etc. Modulation that extends beyond this conventional regime holds the promise to revolutionize all the aforementioned electro-optic physics and technologies. Nevertheless, it has been difficult to explore due to the absence of a consistent theoretical framework for addressing the challenges posed by the parameter regime of $\Omega, \omega_{\mathrm{BW}} \gg \omega_{\mathrm{R}}$. The conventional coupled-mode theory[15] treats cavity EO modulation as a tight-binding model, where frequency modes are coupled to their nearest neighbors under single-tone modulation. However, this framework fails to accurately describe the system when $\Omega \gg \omega_{\mathrm{R}}$.

Another commonly used approach, the transfer matrix model, tracks the evolution of cavity EO modulation in the temporal domain. However, the massive matrix computations required for higher-order drive frequencies ($\omega_{\mathrm{BW}} \gg \omega_{\mathrm{R}}$) significantly increase computational complexity. Moreover, incorporating microwave detuning for advanced functionalities, such as frequency boundary effects[31], further adds to these challenges. As result, the rich dynamics governed by strong-coupling and high-bandwidth cavity EO modulation for pulse-comb synthesis, remain unknown.

Here, we address this challenge and demonstrate cavity electro-optic modulation with arbitrary coupling strength and bandwidth of microwave drive signals, using a general and universal framework. We show that in this framework, EO frequency comb and pulse generation, although typically treated as a linear process[1], manifests unconventional higher-order dynamics. Specifically, in the strong-coupling regime ($\Omega \gg \omega_{\mathrm{R}}$), a single-tone microwave drive can open both short-range and long-range coupling (Fig. 1b, right panel), exciting multiple temporal pulses and reshaping the comb shape from conventional linear triangle shape (dB scale) to oscillating shape (Fig. 1c, Strong drive). This effect is also observed in our recent experiment[32], further highlighting the significance of our proposed modeling framework for modulation in strong-coupling and high-bandwidth regime. Further, in the high-bandwidth regime ($\omega_{\mathrm{BW}} \gg \omega_{\mathrm{R}}$), high-frequency microwave drives can significantly control the spatiotemporal EO pulses properties (Fig. 1c, Infinite bandwidth drive), correlated with the synthetic frequency band structure of the EO-modulated cavity. Compared to the widely studied Kerr combs that rely on $\chi^3$ nonlinearity, where comb and pulse dynamics are typically constrained by a fixed optical pump, the cavity electro-optic comb offers excellent tunability by adjusting the modulation strength and signal bandwidth. We demonstrate that EO pulse-comb synthesis can be fully optimized through machine-learning algorithms. This approach enables a new paradigm for integrating complex EO pulse-comb design with machine learning, offering significant potential for advanced, customizable pulse-comb systems. Our findings show the cavity EO system can have intricate dynamics, while further possessing complementary, active microwave control, enabling machine learning assisted inverse design for arbitrary comb shaping.

## Results

### Framework for strong-coupling and high-bandwidth cavity EO modulation

We start by deriving a Hamiltonian that includes the strong-coupling and high-bandwidth regimes of cavity EO modulation. In conventional descriptions, when a sinusoidal modulation is applied, the Hamiltonian is typically expressed as $H = \sum_n \left[\omega_n a_n^\dagger a_n + g(t)\left(a_{n+1}^\dagger a_n + h.c.\right)\right]$ with $g(t) = \Omega \cos \omega t$ and $\hbar = 1$. The coupling strength $\Omega$ is directly related to the modulation index $\beta$ with $\beta = 2\pi \frac{\Omega}{\omega_{\mathrm{R}}} = \Omega \times t_{\mathrm{R}} = \frac{\Omega}{f_{\mathrm{R}}}$, where $t_R$ is the round-trip time of the cavity (see details in Supplementary Information) and $f_{\mathrm{R}} = \omega_{\mathrm{R}}/2\pi$ is the roundtrip frequency related to the FSR of the cavity. Such a model indicates that coupling occurs only between nearest neighbor energy levels and has been widely adopted in previous works[14,15,28,33–40]. However, this conventional model breaks down when the coupling strength $\Omega$ approaches or even exceeds the FSR of the cavity. We show that to capture the complete dynamics of cavity EO modulation, the coupling term should instead be written as $g(t) = i f_{\mathrm{R}} e^{i \int dz \frac{k}{2n^2} \Delta\varepsilon(z,t)}$, where $n$ and $k$ are the refractive index and wavevector evaluated at the frequency of the pump laser, respectively, $\Delta\varepsilon(z,t) = \varepsilon_0 \chi^{(2)} E_{\omega}(z,t)$ is local dielectric constant oscillation inside the cavity induced by EO modulation, $\chi^{(2)}$ is the second-order nonlinear coefficient, $E_{\omega}(z,t)$ is the modulating microwave field, and the integration is performed across the modulation region. The integration in $g(t)$ abstracts away the specific structure of the modulation electrode in the subsequent discussion and allows us to treat the phase modulation itself as a black box, leading to an interaction Hamiltonian

$$H = \sum_n \left[ \omega_n a_n^{\dagger} a_n + g(t) \sum_m \left( a_{n+m}^{\dagger} a_n + h.c. \right) \right]$$

where the coupling $g(t)$ reduces to

$$g(t) = i f_{\mathrm{R}} \left( e^{i\beta \cos \omega t} - 1 \right) = i f_{\mathrm{R}} \left( e^{i \frac{\Omega}{f_{\mathrm{R}}} \cos \omega t} - 1 \right)$$

in the case of sinusoidal modulation. When $\Omega \ll \omega_{\mathrm{R}}$, our Hamiltonian model naturally reduces to conventional Hamiltonian with $g(t) = \Omega \cos \omega t$.

It is important to note that, in our model, even a single-tone sinusoidal modulation can induce both short-range ($a_{n+1}^{\dagger} a_n$) and long-range ($a_{n+m}^{\dagger} a_n, m > 1$) interactions between different energy levels (or frequency modes) (Fig. 1b). We emphasize that the electro-optic (EO) effect, as a second-order nonlinear process, does not inherently introduce long-range interactions. Instead, such interactions arise from the cavity EO modulation, which operates on the time scale of discrete steps with a step size of $t_{\mathrm{R}}$. It is worth mentioning that both our Hamiltonian model and the conventional Hamiltonian model discussed above are formulated within this same discrete time-step framework. The non-Hermitian property of the Hamiltonian originates from the discrete time-step, therefore does not violate energy conservation by phase modulation. We also note that although there may exist other effects such as $\chi^{(3)}$nonlinearities, thermal effect and dispersion effect in the cavity modulation system, these effects remain the same in weak and strong-coupling regime since the weak/strong coupling is determined by the microwave modulation strength. (See Supplementary Information for further details).

Such a Hamiltonian leads to a set of equations of motion in the discrete time scale, where the discrete time differential $\partial a_n / \partial T \equiv (a_n(t + t_{\mathrm{R}}) - a_n(t))/t_{\mathrm{R}}$, where $T = 1t_{\mathrm{R}}, 2t_{\mathrm{R}}, \cdots$, with $J_m(\beta)$ denotes the Bessel function of order m, and $\kappa = \kappa_i + \kappa_e$, $\kappa_i$, $\kappa_e$ represents the total loss, intrinsic loss and external loss, respectively.

$$\frac{\partial a_n}{\partial T} = \left(f_{\mathrm{R}}(J_0(\beta) - 1) - \frac{\kappa}{2}\right) a_n + f_{\mathrm{R}} \sum_{m \neq 0} i^m J_m(\beta) a_{n+m} + \sqrt{\kappa_e}\sqrt{P_{in}}\delta_{n,0}$$

When $\beta \ll 1$ (i.e. $\Omega \ll \omega_{\mathrm{R}}$), the equation of motions reduce to the conventional weak electro-optic case[5].

$$\frac{\partial a_n}{\partial T} = -\frac{\kappa}{2} a_n + i\left(\frac{\Omega}{2} a_{n+1} + \frac{\Omega}{2} a_{n-1}\right) + \sqrt{\kappa_e}\sqrt{P_{in}}\delta_{n,0}$$

The overall cavity EO modulation beyond the conventional regime ($\Omega, \omega_{\mathrm{BW}} \gg \omega_{\mathrm{R}}$) can be intuitively understood by considering cavity resonant transmission spectrum (or energy levels) experiencing periodic oscillation (with oscillation amplitude $\Omega$ in frequency domain) when modulated by the microwave. Within one period, for weak drive, a single energy level (i.e. the pump resonance) sweeps over the pump signal twice (Fig. 1c, Weak drive), leading to the conventional two-pulse excitation in EO frequency comb. When $\Omega > \omega_{\mathrm{R}}$, the modulation strength $\Omega$ is large enough to modulate the neighbor energy levels/resonances overlapping with the pump signal (Fig. 1c, strong drive), leading to multi-pulses excitation. Furthermore, the generation of pulses is mainly due to the abrupt movement of the resonance which modifies the amplitude of the pump light. As a result, a high-bandwidth signal ($\omega_{\mathrm{BW}} \gg \omega_{\mathrm{R}}$) increases the speed of resonance movement, leading to pulse compression and broadened, flattened comb spectrum (Fig. 1c, High-bandwidth drive, Infinite bandwidth drive).

**Strong-coupling EO modulation-induced unconventional EO comb and pulse synthesis**

Next, we use our model to construct the phase diagram of cavity EO modulation, defined by the coupling strength $\Omega$ and optical pump detuning $\Delta$, containing higher-order pulse-comb dynamics in the strong-coupling regime ($\Omega \gg \omega_{\mathrm{R}}$). This is, in analogous to the phase diagram of the Kerr soliton in which soliton dynamics is governed by the pump power and optical detuning. Different detuning $\Delta$ and modulation strength $\Omega$ determine the pulse number and comb shape within the cavity. The 2-pulse regime corresponds to the conventional EO comb, while multi-pulse (>2) regimes correspond to the strong-coupling regime (Fig. 2a). We show EO comb generation for $\Omega = 0.25 \times \omega_{\mathrm{R}}, 1.75 \times \omega_{\mathrm{R}}, 2.85 \times \omega_{\mathrm{R}}$. The EO comb spectrum changes from a classical triangular shape to a periodic envelope, representing the emergence of long-range jumps between energy levels. Therefore, the temporal pulse number increases from 2 to 6 to 10 for regimes $\Omega < \omega_{\mathrm{R}}$, $\omega_{\mathrm{R}} < \Omega < 2\omega_{\mathrm{R}}$, and $2\omega_{\mathrm{R}} < \Omega < 3\omega_{\mathrm{R}}$, respectively. Note that, before the emergence of new pulses, i.e. when increasing $\Omega$ within a regime of fixed pulse number (as shown in Fig. 2a), the pulse duration is also compressed, which is attributed to the increased rising/falling time of the sinusoidal signal as its amplitude increases, leading to a faster sweeping of resonances over the pump signal.

We further explain that these higher-order dynamics arise from the overlap of energy bands in the synthetic frequency crystal. Cavity EO modulation induces EO coupling among different frequency modes separated by the $f_R$, thus forming a synthetic crystal with lattice coupling created by the EO modulation[34,38]. This leads to multiple independent energy bands in synthetic frequency space with energy bands separated by the $f_R$ and single band range determined by the coupling strength $\Omega$ (Fig. 2c). In the weak-coupling regime ($\Omega < \omega_R$), a single pump field excites two energy states within the single band, corresponding to two pulses (Fig. 2c, Weak-coupling band). When $\Omega > \omega_R$, the band overlap facilitates the excitation of multiple energy states across different bands (Fig. 2c, 1st-order strong-coupling band and 2nd-order strong-coupling band). The number of overlapping bands corresponds exactly to half the number of pulses in the cavity and this band overlap creates state "jumps" between the overlapping bands, which are crucial for long-range, higher-order dynamics. This implies that, in addition to the pump mode in conventional electro-optic modulation, extra cavity modes (both neighbor modes and even next-nearest modes) are simultaneously excited. At this point, the system enters the strong-coupling regime, where conventional electro-optic modulation coupled-mode theory breaks down, and the number of generated electro-optic pulses is twice the number of excited cavity modes. Notably, when $\Omega \ll \omega_R$, a forbidden gap exists between the bands, indicating that the system is insulated from a detuned pump, preventing the EO comb generation in this regime (pump-insulation regime in Fig. 2a). In the strong-coupling regime ($\Omega > \omega_R$), however, the forbidden gap disappears, leading to a phase transition from an insulating state to a fully conducting state for arbitrary pump detuning. We emphasize that, in this regime, the EO modulation system exhibits robustness against pump detuning, enabling excitation of EO comb with arbitrary detuned pump (Fig. 2b, Pump detuning robust EO comb), potentially eliminating the need for a resonant pump for cavity-based EO combs (Fig. 2a).

It is important to note that optical detuning could reduce the threshold for entering the strong-coupling regime. The pump insulation regime connects to the 4-pulse strong-coupling regime at half FSR detuning ($\Delta = \frac{\omega_R}{2}$), indicating that only $\Omega > \frac{\omega_R}{2}$ is required to directly skip the 2-pulse regime and enter the strong-coupling regime. This is due to the boundary between the weak-coupling and strong-coupling regime varies with different pump detuning. Specifically, the threshold condition is $\Omega + |\Delta| > \omega_R$ . At maximum detuning ($\Delta = \pm\omega_R/2$ ), the strong-coupling threshold reduces to the lowest value (Fig. 2e), reducing the voltage requirement for strong EO effects by 50%. This helps mitigate the technical challenges associated with developing microwave optoelectronic devices for achieving strong EO coupling. We scan the pump detuning to calculate the EO conversion efficiency η and gradually increased the modulation strength to $\Omega = \frac{\omega_R}{2}$. The pump isolation regime (where η approaches zero) was gradually compressed until it completely disappeared (Fig. 2d) thus lasers with any detuning can excite the EO comb.

**High-bandwidth EO modulation-induced comb and band structure shaping**

We next extend the modulation to high-bandwidth regime ($\omega_{BW} \gg \omega_R$), enabling the shaping of comb and synthetic band structure using complex modulation signals. This can be achieved by setting $g(t) = i f_R (e^{i\frac{\Omega(t)}{f_R}} - 1)$ with $\Omega(t)$ related to the phase modulation induced by an arbitrary modulation waveform. Notably, we apply the Fourier transform to the function $exp(i\frac{\Omega(t)}{f_R})$, rather than directly to the modulation waveform itself. This results in an additional zero-frequency component, which modulates the loss of pump signal. This can be understood as both complex waveform modulation and strong-coupling introducing additional loss channels for the pump, thereby altering the overall pump loss. We show the frequency spectrum and band structure under square, triangular and ladder waveforms in the strong-coupling regime (Fig. 3a, b), indicating a direct correlation between the modulation waveform and the synthetic band structure. We point out that high-bandwidth microwave driving can induce long-range coupling just as strong-coupling regime can also create long-range coupling. However, we emphasize that one cannot create the same physical dynamics in the strong-coupling regime by adding long-range couplings with multi-tone, weak-coupling drives. There exists no corresponding multi-tone waveform (weak-coupling regime) that can generate the energy band overlap in the strong-coupling case. In the cavity electro-optic modulation parameter space, modulation strength and bandwidth are fully independent degrees of freedom. Consequently, increasing the modulation strength to enter the strong-coupling regime represents a fundamental change in the system's physical behavior. The arbitrary waveform driving enabled band engineering here enables the active and dynamic tuning of band structures in the synthetic frequency dimension by using high-bandwidth EO modulation, offering potential for applications in controlling the dispersion relations of photonic crystals, topological photonics and other related areas.

**Machine-learning-informed arbitrary EO comb inverse design**

Finally, we demonstrate arbitrary EO comb shaping with machine-learning-based microwave signal inverse design (Fig. 4a). A long-standing challenging of cavity EO combs is the spectral linear loss (in dB scale) of the comb lines, with a slope determined by the cavity linewidth $\kappa$ and the electro-optic modulation strength Ω, or $\kappa/\Omega$. By optimizing the $\Omega(t)$ with a given target comb shape using machine learning, we show that the comb slope can be significantly flattened with a high-bandwidth microwave signal ( $\omega_{BW} \gg \omega_R$ ). Furthermore, we show that introducing frequency boundaries using microwave detuning[31,41] can further flatten the comb shape. The detuning-induced frequency boundary typically creates energy reflection in the frequency domain and leads to spectral fluctuations due to spectral interference[31]. Machine learning with high-bandwidth modulation can effectively remove this spectral fluctuation. To this end, we show a microwave signal that can result in a flat cavity electro-optic comb, with $\omega_{\mathrm{R}} = 2\pi \times 3$ GHz, $\omega_{\mathrm{BW}} = 2\pi \times 30$ GHz, and $\delta_{\mathrm{MW}} = 22$ MHz. The flat comb achieves a 3 dB-bandwidth about 200 lines, with a slope of 0.03 dB/line (Fig. 4b, c), representing a tenfold improvement in flatness compared to single-tone electro-optic combs under the same microwave power. We choose the microwave bandwidth $\omega_{\mathrm{BW}} = 2\pi \times 30$ GHz, therefore in actual experiment, one can drive this arbitrary waveform by a 30 GHz arbitrary waveform generator. On the state-of-the-art thin-film lithium niobate platform, a 2 cm electrode for phase modulation reduces the half-wave voltage to approximately 2 V[42], achieving the 16 dBm microwave power threshold required to enter the strong-coupling regime (assuming $\Delta = \omega_{\mathrm{R}}/2$), offered by commercially available microwave amplifiers. Consequently, cavity electro-optic modulation with arbitrarily coupling strengths and bandwidths can be readily implemented in experiments.

Furthermore, machine learning based inverse design can bring additional insight into the cavity EO modulation dynamics. By setting the target comb shape as only one side of the comb compared to the conventional EO comb, we found that the falling and rising edge of the modulation signal directly correlated to the right and left side of the comb spectrum. For example, eliminating the falling edge of the modulation signal can completely remove one side of the comb spectrum without affecting the other side (Fig. 4d, rising and falling edge pulse in different modulation waveform). Inspired by this, we demonstrate arbitrary comb shape design with machine learning algorithms, achieving single-sided-flat comb, one-side-flat with one-side tilt comb, unequal-arm-flat comb, and so on (Fig. 4e, ML-assisted arbitrary spectra generation). We also demonstrate that moving the position of the boundary by varying the microwave detuning may reconfigure the flat cavity EO comb bandwidth (Fig. 4f, Detuning induced programmable bandwidth flat-top comb). These results are enabled by our general framework that is compatible with arbitrary microwave detuning, strength, and bandwidth. We note that optimization algorithms based on conventional models, especially the transfer matrix approach, creates significant computational complexity. The transfer matrix approach calculates the time domain pulse and needs to do the Fourier transform which scales as $O(N\log N)$, while the coupled mode approach based on the effective Hamiltonian directly solves the coupled mode equations and is compatible with the GPU accelerators and time complexity scales as $O(N)$ owing to its sparse matrix property, when detuning is introduced in the system, especially in the high-bandwidth modulation regime, the transfer matrix model complexity further increases.

## Discussion

In summary, we demonstrate a general framework for strong-coupling, high-bandwidth cavity EO modulation, featuring unconventional high-order dynamics with multiple pulses and arbitrary spectral shaping capability that is fully compatible with machine learning optimization. We expand the parameter space of cavity EO systems across three dimensions: coupling strength $\Omega$, modulation bandwidth BW (microwave domain), and pump detuning $\Delta$ (optical domain), offering deeper insights into EO comb generation and coupling mechanisms. This breaks the limit of conventional EO modulation in weak modulation strength, narrow bandwidth, and small detuning. Such a framework agrees well with our experimental observations[32] and opens the door for the combination of cavity EO comb and pulse optimization using machine learning. The spectral flattening utilizing high-bandwidth modulation and frequency boundaries presents a way to address the fundamental challenge of cavity EO comb slope. The introduction of microwave detuning to flatten the comb spectrum leverages the power of physics-informed machine learning model, achieving a substantial reduction in the computational resources needed for optimizing this complex photonic inverse design problem. Furthermore, the pump wavelength robustness to detuning in the strong EO regime breaks the conventional requirement for precise pump wavelength matching, removing the critical bottleneck of the cavity EO comb and further advancing the comb reconfigurability for tunable and multi-color pumped EO combs.

The modeled high-order dynamics with long-range coupling have a significant impact on current cavity EO modulation-based applications such as topological photonics[13–15] and non-Hermitian physics[43–45]. The established correspondence between the synthetic band structure and modulation waveforms opens the possibility for actively controlled topological non-Hermitian band structures as well as synthetic dispersion control for slow light. The high-bandwidth modulation also unlocks the potential for programmable temporal control on the position, intensity, and width of cavity EO pulses, with significant implications for ultrafast applications including octave-spanning frequency combs[46,47], optical parametric oscillations[48–50], and Raman scattering spectroscopies[51–53]. The advantages demonstrated on machine-learning-based spectral control for broadening, flattening, and arbitrary shaping overcome the flexibility limitations of the comb spectrum that exists in almost all types of frequency combs including EO combs[4,5], Kerr combs[54–58], and mode lock lasers[59]. This unique property makes strong-coupling, high-bandwidth cavity EO combs a promising source for optical communications[9,60–63], photonic computing[64–68], and microwave photonics[11,12,69–73]. It also opens the door for study the quantum dynamics of high-order dynamics of EO modulation for microwave-to-optical transducer[74–76] and microwave entanglement[77,78]. Our work opens an underexplored regime for cavity EO modulation, paving the way for machine-learning-based, programmable, strong-coupling electro-optics applications.

**Materials and methods:** Supplementary Information is available for this paper, providing details of the theoretical model, additional data, extended discussion for cavity electro-optic modulation and machine learning based flat comb generation methodologies.

**Acknowledgements:** T.L. thanks Chenyang Cao, Xianpeng Lv, Xiongfeng Zhang, Junting Bie, Yiming Lei, Yingchuan Qi, and Ruokai Zheng for discussions.

**Author contributions:** Y.H., M.L., T.L., Y.S. conceived the project. T.L. developed modeling and performed numerical simulation with Y.S. assisting. T.L., Y.S., Y.X., M.L., and Y.H. interpreted and analyzed the data. T.L. and Y.H. wrote the manuscript with contributions from all authors. Y.H., M.L., and Q.G. supervised the project.

**Funding:** National Research Foundation funded by the Korea government, (NRF-2022M3K4A1094782).

**Data availability**: The data that support the plots within this paper and other findings of this study are available from the corresponding authors upon reasonable request.

**Code availability**: The code used to produce the plots within this paper is available from the corresponding authors upon reasonable request

**Conflict of interests:** M.L. is involved in developing lithium niobate technologies at HyperLight Corporation.

**References**

1. Parriaux, A., Hammani, K. & Millot, G. Electro-optic frequency combs. *Adv. Opt. Photonics* **12**, 223 (2020).

2. Carlson, D. R. *et al.* Ultrafast electro-optic light with subcycle control. *Science* **361**, 1358–1363 (2018).

3. Ho, K.-P. & Kahn, J. M. Optical frequency comb generator using phase modulation in amplified circulating loop. *IEEE Photonics Technol. Lett.* **5**, 721–725 (1993).

4. Zhang, M. *et al.* Broadband electro-optic frequency comb generation in a lithium niobate microring resonator. *Nature* **568**, 373–377 (2019).

5. Hu, Y. *et al.* High-efficiency and broadband on-chip electro-optic frequency comb generators. *Nat. Photonics* **16**, 679–685 (2022).

6. Yang, Q.-F., Hu, Y., Torres-Company, V. & Vahala, K. Efficient microresonator frequency combs. *eLight* 18 (2024).

7. Kikuchi, K. Fundamentals of Coherent Optical Fiber Communications. *J. Light. Technol.* **34**, 157–179 (2016).

8. Marin-Palomo, P. *et al.* Microresonator-based solitons for massively parallel coherent optical communications. *Nature* **546**, 274–279 (2017).

9. Lukens, J. M. *et al.* All-Optical Frequency Processor for Networking Applications. *J. Light. Technol.* **38**, 1678–1687 (2020).

10. Yoo, S. J. B. Wavelength conversion technologies for WDM network applications. *J. Light. Technol.* **14**, 955–966 (1996).

11. Xu, X. *et al.* 11 TOPS photonic convolutional accelerator for optical neural networks. *Nature* **589**, 44–51 (2021).

12. Feldmann, J. *et al.* Parallel convolutional processing using an integrated photonic tensor core. *Nature* **589**, 52–58 (2021).

13. Wang, K., Dutt, A., Wojcik, C. C. & Fan, S. Topological complex-energy braiding of non-Hermitian bands. *Nature* **598**, 59–64 (2021).

14. Wang, K. *et al.* Generating arbitrary topological windings of a non-Hermitian band. *Science* **371**, 1240–1245 (2021).

15. Dutt, A. *et al.* A single photonic cavity with two independent physical synthetic dimensions. *Science* **367**, 59–64 (2020).

16. Yuan, L., Lin, Q., Xiao, M. & Fan, S. Synthetic Dimension in Photonics. *Optica* **5**, 1396 (2018).

17. Chang, L., Liu, S. & Bowers, J. E. Integrated optical frequency comb technologies. *Nat. Photonics* **16**, 95–108 (2022).

18. Diddams, S. A., Vahala, K. & Udem, T. Optical frequency combs: Coherently uniting the electromagnetic spectrum. *Science* **369**, eaay3676 (2020).

19. Zhang, J. *et al.* Quantum internet using code division multiple access. *Sci. Rep.* **3**, 2211 (2013).

20. Kues, M. *et al.* Quantum optical microcombs. *Nat. Photonics* **13**, 170–179 (2019).

21. Lu, H.-H., Liscidini, M., Gaeta, A. L., Weiner, A. M. & Lukens, J. M. Frequency-bin photonic quantum information. *Optica* **10**, 1655 (2023).

22. Menicucci, N. C., Flammia, S. T. & Pfister, O. One-Way Quantum Computing in the Optical Frequency Comb. *Phys. Rev. Lett.* **101**, 130501 (2008).

23. Hu, Y. *et al.* Integrated electro-optics on thin-film lithium niobate. *Nat. Rev. Phys.* **7**, 237–254 (2025).

24. Karpov, M. *et al.* Dynamics of soliton crystals in optical microresonators. *Nat. Phys.* **15**, 1071–1077 (2019).

25. Cole, D. C., Lamb, E. S., Del'Haye, P., Diddams, S. A. & Papp, S. B. Soliton crystals in Kerr resonators. *Nat. Photonics* **11**, 671–676 (2017).

26. Li, G. *et al.* Ring Resonator Modulators in Silicon for Interchip Photonic Links. *IEEE J. Sel. Top. Quantum Electron.* **19**, 95–113 (2013).

27. Li, G. *et al.* 25Gb/s 1V-driving CMOS ring modulator with integrated thermal tuning. *Opt. Express* **19**, 20435 (2011).

28. Hu, Y. *et al.* On-chip electro-optic frequency shifters and beam splitters. *Nature* **599**, 587–593 (2021).

29. Zhang, M. *et al.* Electronically programmable photonic molecule. *Nat. Photonics* **13**, 36–40 (2019).

30. Herrmann, J. F. *et al.* Mirror symmetric on-chip frequency circulation of light. *Nat. Photonics* **16**, 603–608 (2022).

31. Hu, Y. *et al.* Mirror-induced reflection in the frequency domain. *Nat. Commun.* **13**, 6293 (2022).

32. Yunxiang Song *et al.* Universal dynamics and microwave control of programmable cavity electro- optic frequency combs. Preprint at https://arxiv.org/abs/2507.21835 (2025).

33. Dutt, A. *et al.* Experimental band structure spectroscopy along a synthetic dimension. *Nat. Commun.* **10**, 3122 (2019).

34. Hu, Y., Reimer, C., Shams-Ansari, A., Zhang, M. & Loncar, M. Realization of high-dimensional frequency crystals in electro-optic microcombs. *Optica* **7**, 1189 (2020).

35. Lustig, E. *et al.* Photonic topological insulator in synthetic dimensions. *Nature* **567**, 356–360 (2019).

36. Senanian, A., Wright, L. G., Wade, P. F., Doyle, H. K. & McMahon, P. L. Programmable large-scale simulation of bosonic transport in optical synthetic frequency lattices. *Nat. Phys.* **19**, 1333–1339 (2023).

37. Balčytis, A. *et al.* Synthetic dimension band structures on a Si CMOS photonic platform. *Sci. Adv.* **8**, eabk0468 (2022).

38. Ozawa, T., Price, H. M., Goldman, N., Zilberberg, O. & Carusotto, I. Synthetic dimensions in integrated photonics: From optical isolation to four-dimensional quantum Hall physics. *Phys. Rev. A* **93**, 043827 (2016).

39. Lustig, E. & Segev, M. Topological photonics in synthetic dimensions. *Adv. Opt. Photonics* **13**, 426 (2021).

40. Dutt, A. *et al.* Creating boundaries along a synthetic frequency dimension. *Nat. Commun.* **13**, 3377 (2022).

41. Buscaino, B., Zhang, M., Loncar, M. & Kahn, J. M. Design of Efficient Resonator-Enhanced Electro-Optic Frequency Comb Generators. *J. Light. Technol.* **38**, 1400–1413 (2020).

42. Zhu, D. *et al.* Spectral control of nonclassical light pulses using an integrated thin-film lithium niobate modulator. *Light Sci. Appl.* **11**, 327 (2022).

43. Peng, B. *et al.* Parity–time-symmetric whispering-gallery microcavities. *Nat. Phys.* **10**, 394–398 (2014).

44. Song, W. *et al.* Breakup and Recovery of Topological Zero Modes in Finite Non-Hermitian Optical Lattices. *Phys. Rev. Lett.* **123**, 165701 (2019).

45. Li, G. *et al.* Direct extraction of topological Zak phase with the synthetic dimension. *Light Sci. Appl.* **12**, 81 (2023).

46. Song, Y., Hu, Y., Zhu, X., Yang, K. & Lončar, M. Octave-spanning Kerr soliton frequency combs in dispersion- and dissipation-engineered lithium niobate microresonators. *Light Sci. Appl.* **13**, 225 (2024).

47. Kippenberg, T. J., Holzwarth, R. & Diddams, S. A. Microresonator-Based Optical Frequency Combs. *Science* **332**, 555–559 (2011).

48. Ye, H. *et al.* Electro-optic comb pumped optical parametric oscillator with flexible repetition rate at GHz level. *Opt. Lett.* **46**, 1652 (2021).

49. Wang, Y. *et al.* Toward ultimate-efficiency frequency conversion in nonlinear optical microresonators. *C E N C E V N C E S* (2025).

50. Lin, Y. *et al.* Efficient second-harmonic generation of quasi-bound states in the continuum in lithium niobate thin film enhanced by Bloch surface waves. *Nanophotonics* **13**, 2335–2348 (2024).

51. Butler, H. J. *et al.* Using Raman spectroscopy to characterize biological materials. *Nat. Protoc.* **11**, 664–687 (2016).

52. Ozeki, Y. *et al.* High-speed molecular spectral imaging of tissue with stimulated Raman scattering. *Nat. Photonics* **6**, 845–851 (2012).

53. Wan, S. *et al.* Self-locked broadband Raman-electro-optic microcomb. *Nat. Commun.* **16**, 4829 (2025).

54. Del'Haye, P. *et al.* Optical frequency comb generation from a monolithic microresonator. *Nature* **450**, 1214–1217 (2007).

55. Levy, J. S. *et al.* CMOS-compatible multiple-wavelength oscillator for on-chip optical interconnects. *Nat. Photonics* **4**, 37–40 (2010).

56. Sun, S. *et al.* Integrated optical frequency division for microwave and mmWave generation. *Nature* **627**, 540–545 (2024).

57. Long, J. *et al.* A chip-based optoelectronic-oscillator frequency comb. *eLight* **5**, (2025).

58. Nie, B. *et al.* Soliton microcombs in X-cut LiNbO3 microresonators. *eLight* (2025).

59. Guo, Q. *et al.* Ultrafast mode-locked laser in nanophotonic lithium niobate. *Science* **382**, 708–713 (2023).

60. Wen, H. *et al.* Few-mode fibre-optic microwave photonic links. *Light Sci. Appl.* **6**, e17021–e17021 (2017).

61. Yang, K. Y. *et al.* Multi-dimensional data transmission using inverse-designed silicon photonics and microcombs. *Nat. Commun.* **13**, 7862 (2022).

62. Wang, N. *et al.* Laser$^2$ : A two-domain photon-phonon laser. *Sci. Adv.* **9**, eadg7841 (2023).

63. Liu, H. *et al.* Ultra-compact lithium niobate photonic chip for high-capacity and energy-efficient wavelength-division-multiplexing transmitters. *Light Adv. Manuf.* **4**, 1 (2023).

64. Song, Y. *et al.* Integrated electro-optic digital-to-analog link for efficient computing and arbitrary waveform generation. Preprint at https://arxiv.org/abs/2411.04395 (2024).

65. Hu, Y. *et al.* Integrated lithium niobate photonic computing circuit based on efficient and high-speed electro-optic conversion. Preprint at https://arxiv.org/abs/2411.02734 (2024).

66. Zheng, Y. *et al.* Electro-optically programmable photonic circuits enabled by wafer-scale integration on thin-film lithium niobate. *Phys. Rev. Res.* **5**, 033206 (2023).

67. Lin, Z. *et al.* 120 GOPS Photonic tensor core in thin-film lithium niobate for inference and in situ training. *Nat. Commun.* **15**, 9081 (2024).

68. Zheng, Y. *et al.* Multichip multidimensional quantum networks with entanglement retrievability. *Science* **381**, 221–226 (2023).

69. Li, J., Yi, X., Lee, H., Diddams, S. A. & Vahala, K. J. Electro-optical frequency division and stable microwave synthesis. *Science* **345**, 309–313 (2014).

70. Newman, Z. L. *et al.* Architecture for the photonic integration of an optical atomic clock. *Optica* **6**, 680 (2019).

71. Feng, H. *et al.* Integrated lithium niobate microwave photonic processing engine. *Nature* **627**, 80–87 (2024).

72. Marpaung, D., Yao, J. & Capmany, J. Integrated microwave photonics. *Nat. Photonics* **13**, 80–90 (2019).

73. Bai, B. *et al.* Microcomb-based integrated photonic processing unit. *Nat. Commun.* **14**, 66 (2023).

74. Holzgrafe, J. *et al.* Cavity electro-optics in thin-film lithium niobate for efficient microwave-to-optical transduction. *Optica* **7**, 1714 (2020).

75. McKenna, T. P. *et al.* Cryogenic microwave-to-optical conversion using a triply resonant lithium-niobate-on-sapphire transducer. *Optica* **7**, 1737 (2020).

76. Han, X., Fu, W., Zou, C.-L., Jiang, L. & Tang, H. X. Microwave-optical quantum frequency conversion. *Optica* **8**, 1050 (2021).

77. Warner, H. K. *et al.* Coherent control of a superconducting qubit using light. *Nat. Phys.* **21**, 831–838 (2025).

78. Youssefi, A. *et al.* A cryogenic electro-optic interconnect for superconducting devices. *Nat. Electron.* **4**, 326–332 (2021).

**a**

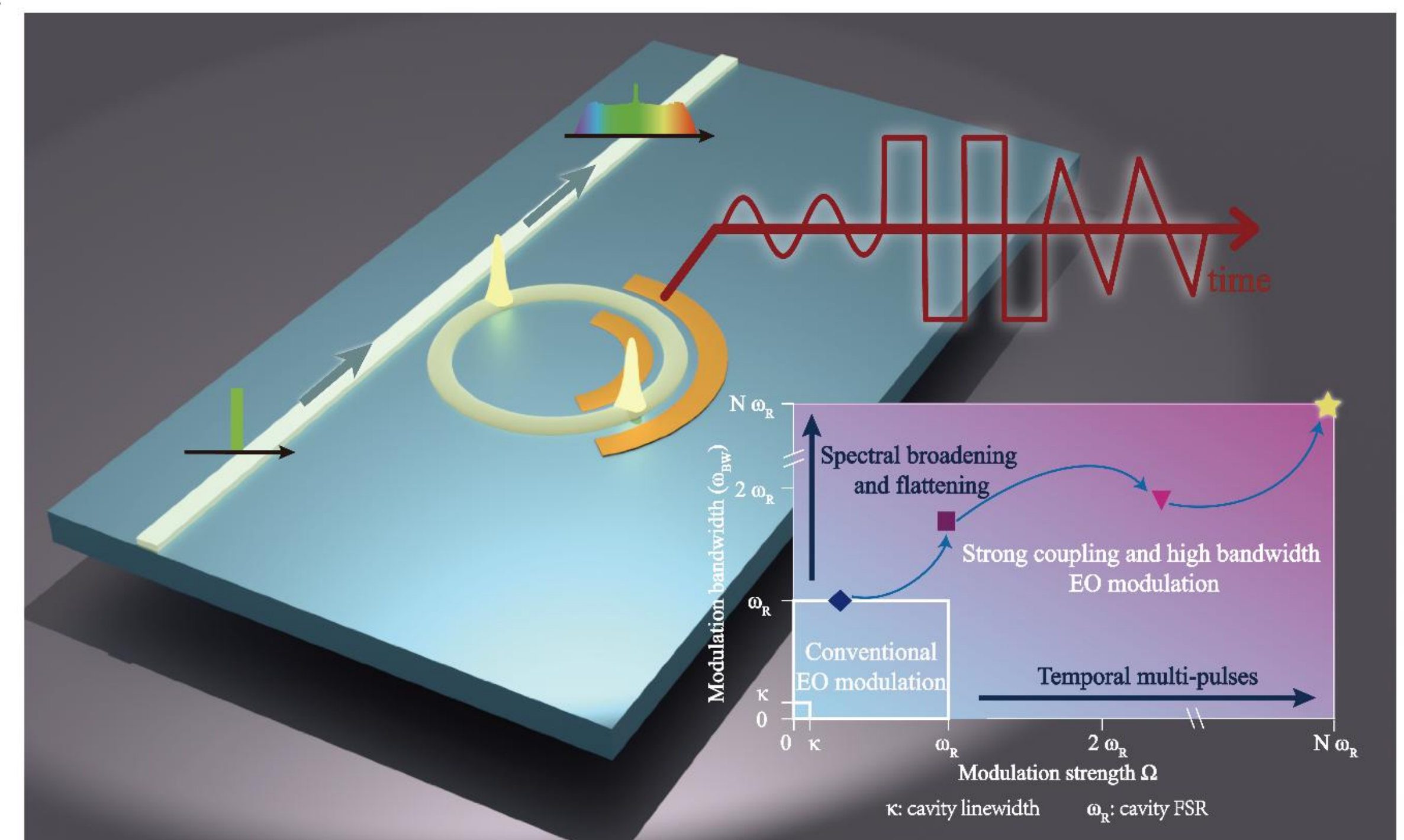

**b**

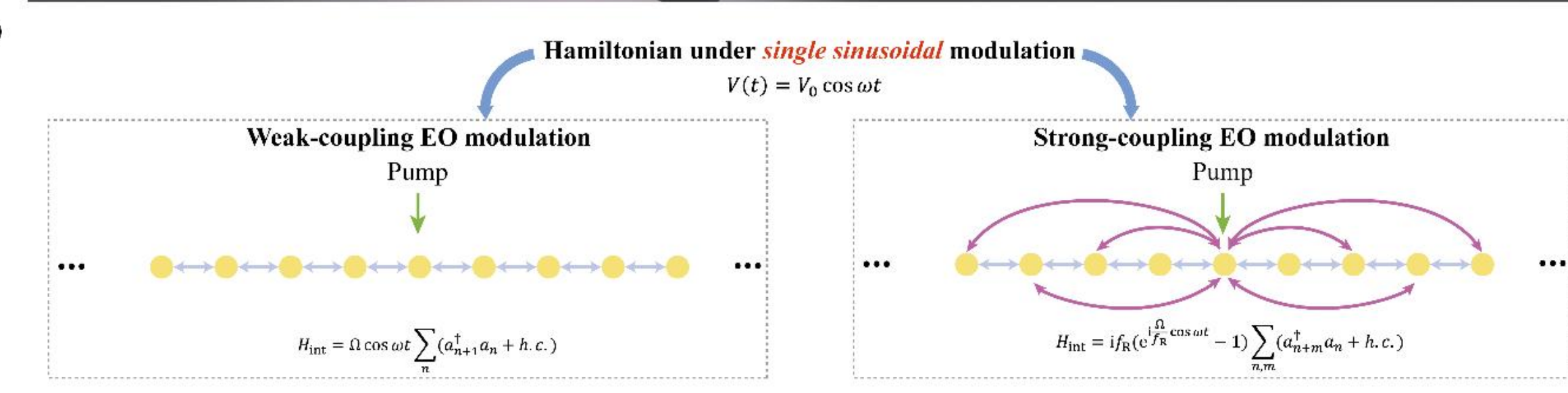

**c**

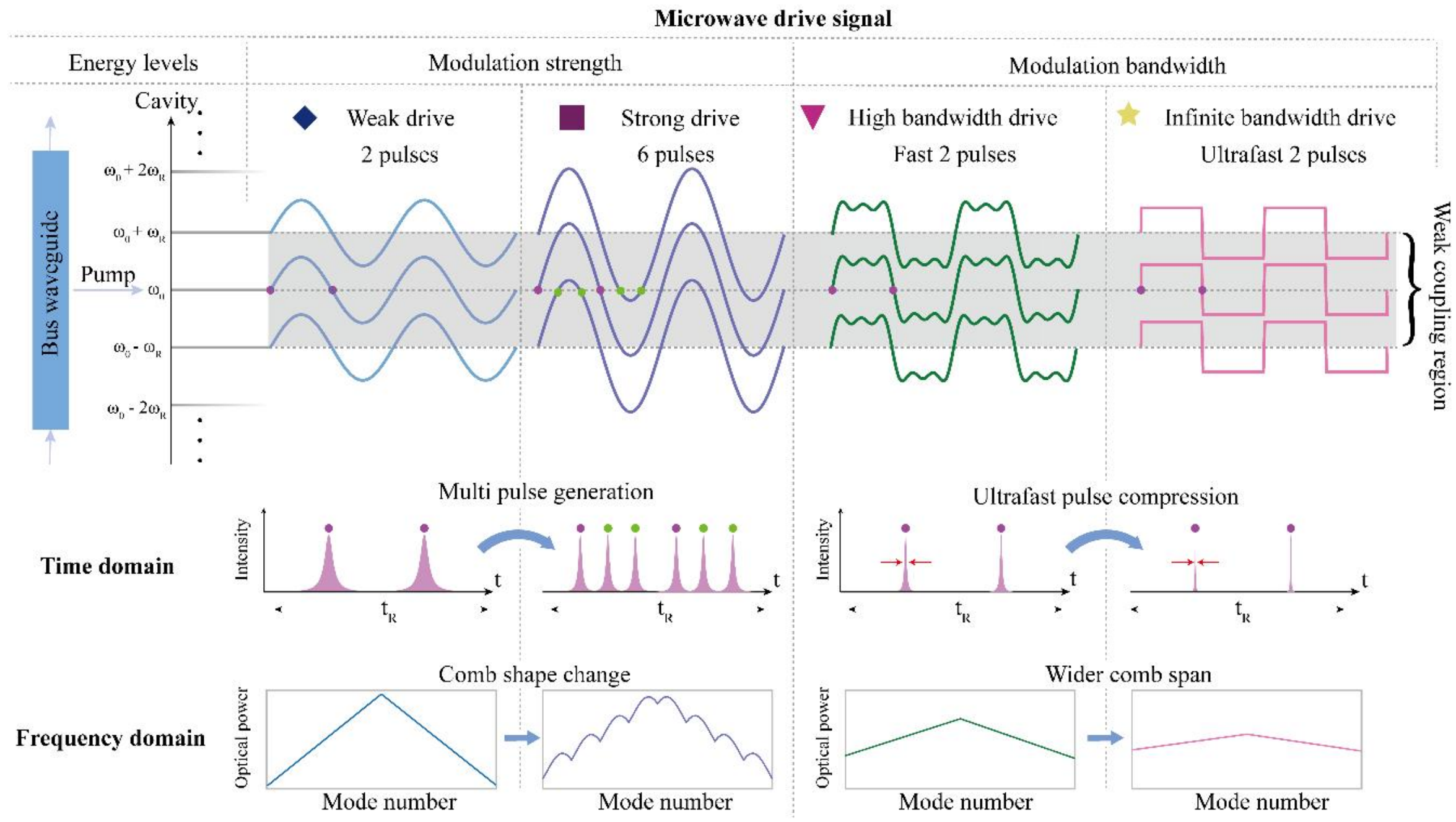

**Figure 1 | Strong-coupling and high-bandwidth cavity EO modulation. a,** Schematic of cavity EO modulation with arbitrary modulation strength and bandwidth (i.e., arbitrary waveforms). The transition from conventional EO modulation regime to the strong-coupling and high-bandwidth regime unveils the unique physical properties of the cavity EO system. **b,** Schematic of the coupling between the frequency lattice points that is induced by conventional and strong-coupling cavity EO modulation. The frequency lattice corresponds to the cavity modes that is separated by the FSR. In the strong-coupling regime, long-range coupling is induced even if only a single-tone modulation with frequency matching the FSR is applied. **c,** Physical picture of arbitrary modulation strength and bandwidth of cavity EO modulation for pulse-comb synthesis. The pump coupled from the bus waveguide excites the cavity center frequency $\omega_0$ and EO modulation couples $\omega_0$ to other resonant levels in the microcavity. Under weak drives, the pump is excited twice in one modulation period, corresponding to two pulses in the cavity. Under strong drive, additional energy levels are excited in one modulation period, increasing the number of pulses in the cavity to six and introducing long-range coupling between different cavity energy levels. Higher modulation bandwidth further allows for waveforms beyond simple sinusoidal modulation, providing more controllable freedom over the pulse widths and coupling between energy levels.

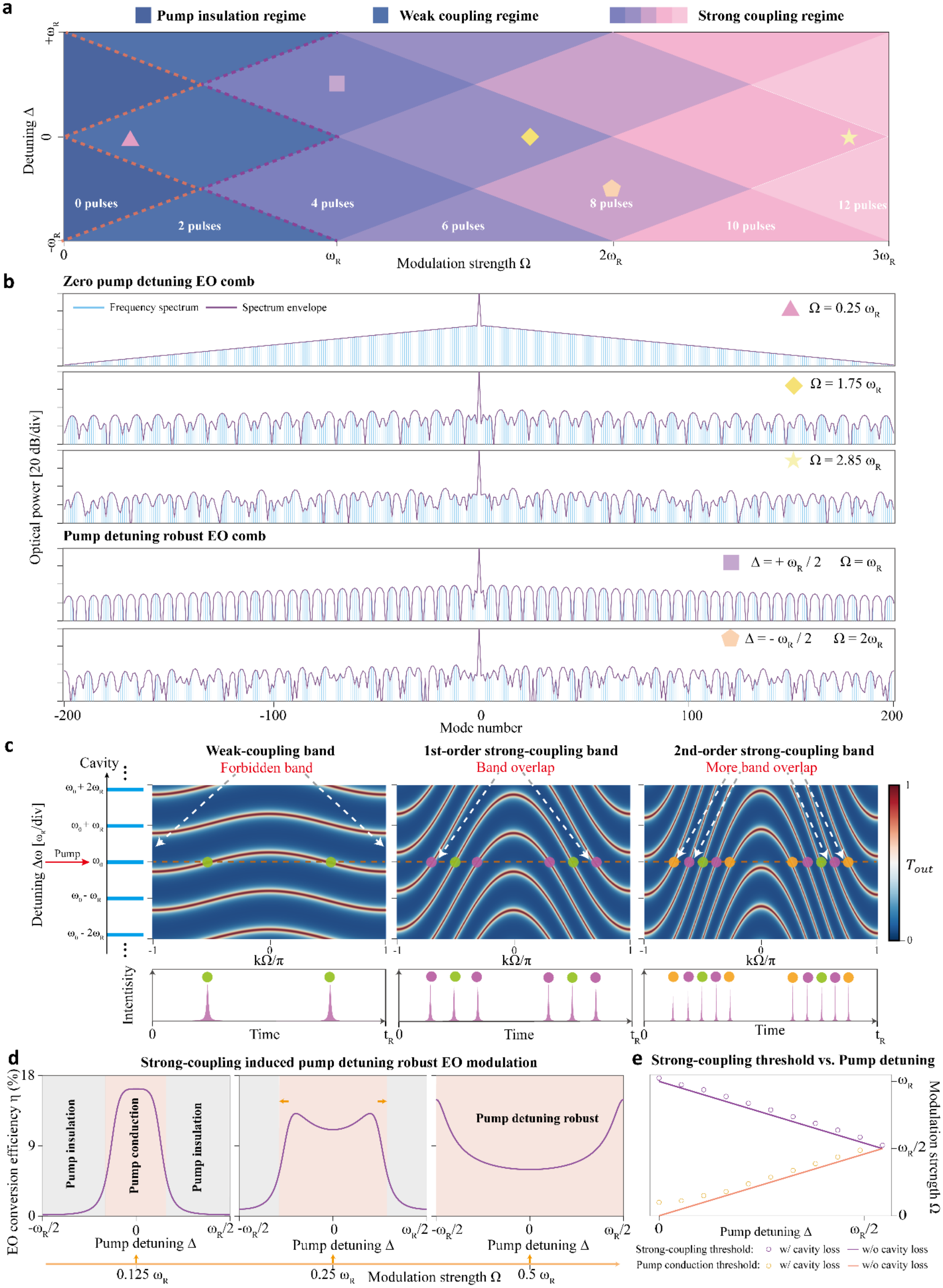

**Figure 2 | Higher-order dynamics of cavity EO comb and pulse synthesis in strong-coupling regime. a,** The phase diagram of EO modulation in the phase space of pump detuning and modulation strength, divided into three regimes: Pump insulation regime, where the pump laser cannot excite electro-optic modulation pulses; Weak coupling regime, where the pump laser excites conventional two pulses per modulation period; and Strong-coupling regime, where the pump laser excites multiple pulses per modulation period. **b,** Zero pump detuning EO comb: the EO modulation spectra correspond to the colored polygons in **a**, with modulation strengths $\Omega = 0.25\omega_\mathrm{R}$, $\Omega = 1.75\omega_\mathrm{R}$, $\Omega = 2.85\omega_\mathrm{R}$ under zero laser detuning $\Delta = 0$. When the modulation strength exceeds $\omega_\mathrm{R}$, the spectra exhibit periodic spectral envelopes; when surpassing $2\omega_\mathrm{R}$, more complex long-range interactions emerge in spectra. Pump detuning robust EO comb: the EO comb spectra corresponding to the colored polygons in **a**, with modulation strength $\Omega = \omega_\mathrm{R}$ and $\Omega = 2\omega_\mathrm{R}$ and the maximum pump detuning $\Delta = \omega_\mathrm{R}/2$ or $\Delta = -\omega_\mathrm{R}/2$, demonstrating the robustness of the cavity EO system to pump detuning under strong-coupling conditions. **c,** The band structures for the three modulation strengths of the zero pump detuning EO comb in panel **b**. Weak coupling band, no band overlapping: there exists forbidden band between different levels which prevents detuned pump exciting EO pulse. 1st-order strong-coupling band: energy bands overlapping appear when the modulation strength steps into strong-coupling regime $\omega_\mathrm{R}$. 2nd-order strong-coupling band: additional overlaps emerge beyond $2\omega_\mathrm{R}$. The number of time-domain electro-optic pulses corresponds to the three modulation strengths in panel **b.** $T_\mathrm{out}$: transmission at the through port of the coupled waveguide. **d,** Increasing the modulation strength of the cavity EO modulation system suppresses the pump insulation regime and creates a fully pump conducting regime under arbitrary pump detuning. The light gray area: pump insulation regime, in which the pump light is not able to excite the generation of EO comb. The dark gray area: pump conduction regime, in which the pump light can excite EO comb. When increasing the modulation strength $\Omega$ to $0.5\omega_\mathrm{R}$, the pump insulation regime disappears and the EO modulation system becomes robust to pump detuning. **e,** Strong-coupling threshold versus pump detuning. The threshold condition becomes $\Omega + |\Delta| > \omega_\mathrm{R}$ and the slight offset between the simulated and theoretical results originates from the cavity's finite loss. When the maximum detuning is reached, the strong-coupling threshold is the lowest, corresponding to a modulation strength of $\Omega = \omega_\mathrm{R}/2$.

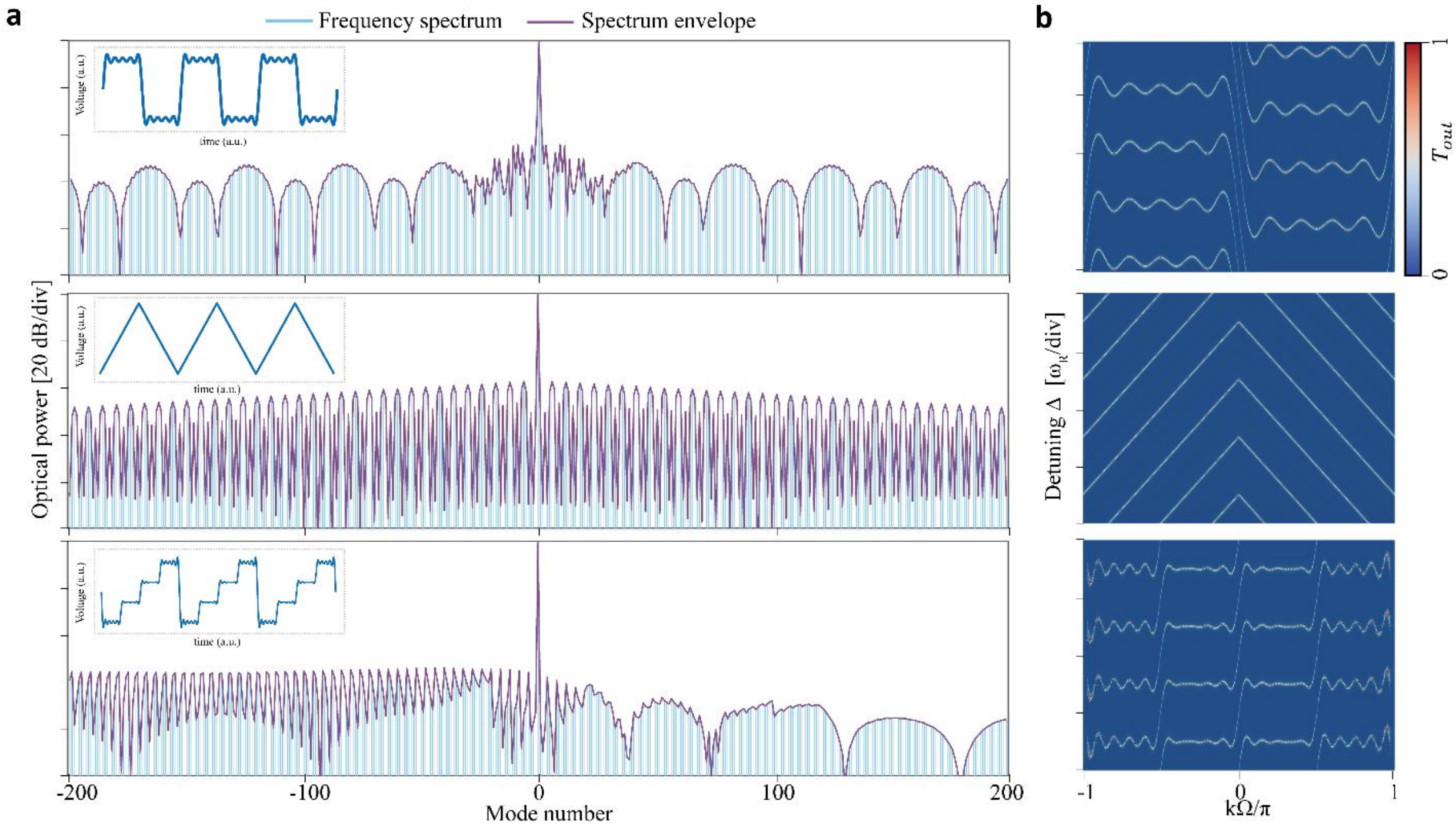

**Figure 3 | Dynamically programable EO frequency comb spectrum with arbitrary driving waveform in high bandwidth regime. a,** Electro-optic modulation spectra under strong-coupling for square wave, triangular wave, and ladder wave modulation. The periodic spectrum envelope reflects the long-range interactions induced by the strong-coupling. The insets within the dashed boxes show the corresponding modulation waveforms. **b,** The band structures of the synthetic frequency lattice under square wave, triangular wave, and ladder wave modulation in panel **a**, indicating a one-to-one correspondence between the modulation waveforms and the frequency lattice band structures.

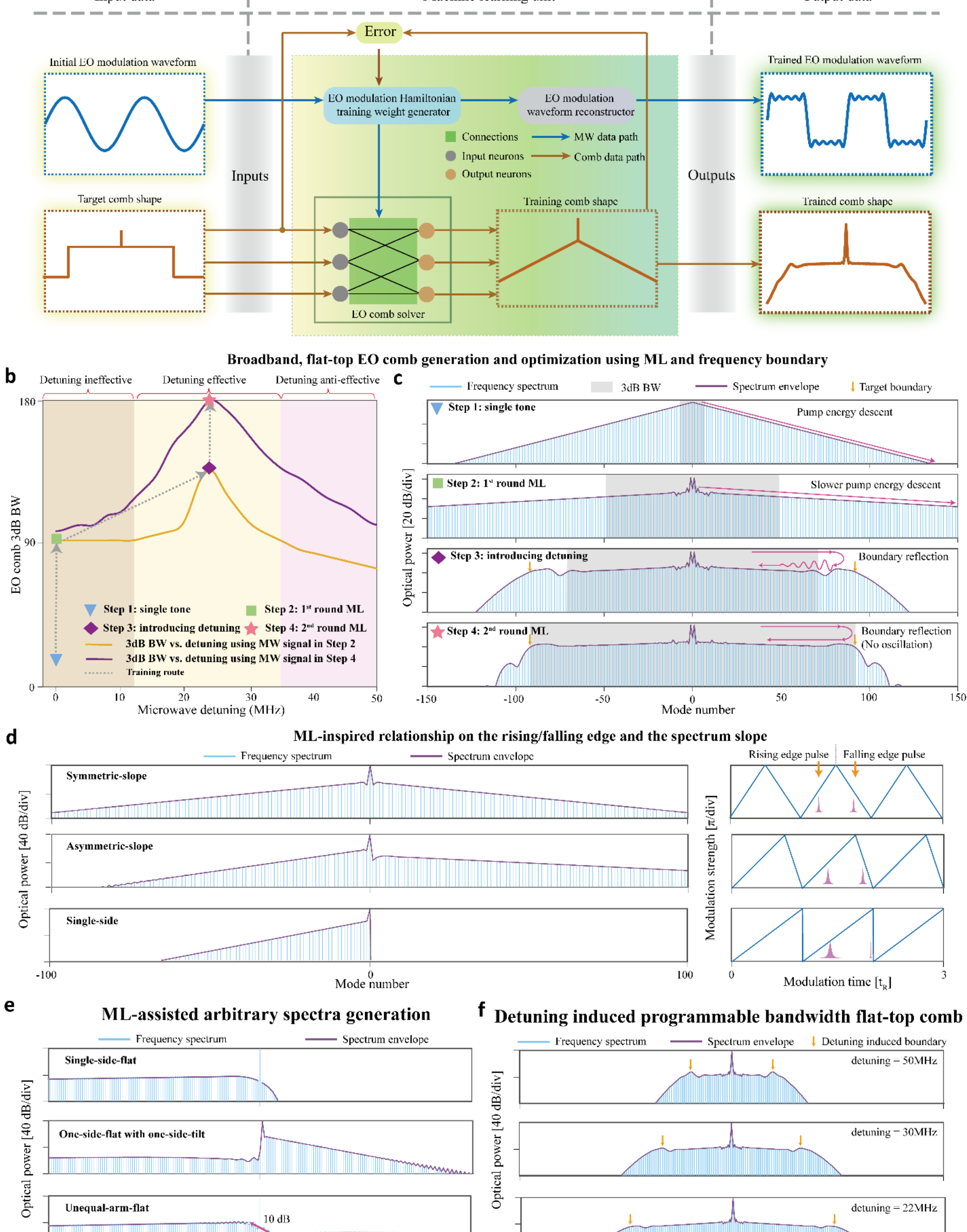

**Figure 4 | Machine-learning-enabled microwave signal inverse design for arbitrary comb shaping with frequency boundary. a,** Machine learning (ML) assisted spectral inverse design of EO comb (e.g., flat-top cavity EO comb). The target comb shape (orange line) and the initial electro-optic modulation waveform (blue line) serve as inputs to the ML unit. The ML process learns the relationship between the microwave drive and comb shape, outputting the trained comb shape with optimized microwave drive signal. **b**, Broadband, flat-top EO comb generation and optimization using ML and frequency boundary. The figure shows the EO comb 3-dB bandwidth versus the microwave detuning (orange and purple traces) with the dashed grey line representing the training route. Four key steps are labeled by triangle, square, diamond, and star, representing the steps for single tone, 1$^{st}$ round ML, introduce detuning, and 2$^{nd}$ round ML, respectively. The introduction of microwave detuning combined with ML can double the 3 dB bandwidth (BW) (from 90 to 180 comb lines). In the above training process, to ensure fairness in comparison, the total microwave powers used in the different training waveforms are the same. Detuning effective regime: boundary effectively localizes pump energy within the preset 3 dB BW. Detuning anti-effective regime: boundary-induced reflection is too strong. Detuning ineffective regime: boundary-induced reflection is too weak. **c,** Comb spectra for steps 1-4. The purple arrows indicate the pump energy flow along the frequency lattice. Introducing microwave detuning boundary can effectively localize the energy within the target BW (orange arrows). The 2$^{nd}$ round of ML (step 4) eliminates the spectral oscillations near the frequency boundary (purple arrows), ultimately generating an ultra-wide, flat-top EO comb. **d,** ML-inspired relationship on the rising/falling edge and the spectrum slope. Reducing the falling edge pulse width decreases the slope of one branch of the EO comb and lowers every comb line power to ensure the energy conservation during the EO comb synthesis. In the extreme case, the pulse width approaches zero and one side of the EO comb disappears. **e**, ML-assisted arbitrary spectra generation of single-side-flat, one-side-flat with one-side tilt, and unequal-arm-flat EO combs. **f**, Flat-top comb bandwidth programming by only adjusting the microwave detuning, without the need of ML retraining.